# CARBON NANOTUBE BASED DELAY MODEL FOR HIGH SPEED ENERGY EFFICIENT ON CHIP DATA TRANSMISSION USING: CURRENT MODE TECHNIQUE

Sunil Jadav[1], Munish Vashistah[2], Rajeevan Chandel[3]

[1,2]Electronics Engineering Department, YMCAUST, Faridabad, Haryana India-121006
[3]Electronics & Communication Engineering, Department, N.I.T, Hamirpur India- 177005

*Abstract:* Speed is a major concern for high density VLSI networks. In this paper the closed form delay model for current mode signalling in VLSI interconnects has been proposed with resistive load termination. RLC interconnect line is modelled using characteristic impedance of transmission line and inductive effect. The inductive effect is dominant at lower technology node is modelled into an equivalent resistance. In this model first order transfer function is designed using finite difference equation, and by applying the boundary conditions at the source and load termination. It has been observed that the dominant pole determines system response and delay in the proposed model. Using CNIA tool (carbon nanotube interconnect analyzer) the interconnect line parameters has been estimated at 45nm technology node. The novel proposed current mode model superiority has been validated for CNT type of material. It superiority factor remains to 66.66% as compared to voltage mode signalling. And current mode dissipates 0.015pJ energy where as VM consume 0.045pJ for a single bit transmission across the interconnect over CNT material. Secondly the damping factor of a lumped RLC circuit is shown to be a useful figure of merit.

## I. INTRODUCTION

Interconnects are one of the essential parts of the VLSI chips. As technology scales down, device dimensions decreases but at the same time chip dimensions increases in order to embed more and more devices on the same chip. As a result, global interconnects causes major delays in the circuits. At deep sub micron technologies these delays are even more than the gate delays and hence need to be reduced [1]. Various techniques to improves the performance of interconnects has been proposed [2]. Repeater insertions method has been suggested by many researchers [3, 4]. But there is some practical limitations to the performance improvement [5]. Moreover repeaters need to be proper sized and should be fixed at proper intervals to achieve optimum results. As an alternative approach to improve the performance of interconnects, current mode signaling has been proposed [6, 7, 8, 9,10,11,12 & 13].

A closed- form RC model for current mode interconnects has been derived using first order moment approximation and boundary condition matching in Ref. [14]. However, as system requirements push for the use of wider low resistance line, the inductance become increasingly dominant under fast transitions in GHz frequency range. In this case a RC delay model in [14] results in an error more than 20% compared to HSPICE simulations when operating in inductance dominated regions. But this aspect is important for current mode interconnects and therefore has been attempted in the present research work. An RLC interconnect line needs to be approximated as a RC line model using inductance-resistance equivalent model. This helps in mitigating the estimated error in [14, 15, and 16] for GHz frequency range. Due to this the overall system performance gets improved in terms of speed, throughput, and energy consumption during transmission of single bit and accuracy at GHz range.

The rest of the paper is organized as follows. In section II inductance equivalent resistance concept is discussed. In section III the proposed problem is defined with analytical model and mathematical formulation is presented for resistive load termination and damping factor is considered for accuracy. Results and their discussion are presented in section IV. Finally conclusions are drawn in section V.

## II. INDUCTANCE-RESISTANCE EQUIVALENT MODEL

The current mode interconnect delay expression is derived through two main steps namely,

(i) Absorbing the line inductance into effective resistance.

(ii) Using transfer function Laplace operator approach and by applying boundary conditions at source and load end of line.

The line inductance is converted into an effective resistance. In case of *RC* interconnects, the equivalent line resistance is "$R_S + R_T$". However, when inductive effect is dominant the equivalent resistance equals "$R_T + 0.65R_S + 0.36Z_0$" where the factors 0.65 and 0.36 reflect the shielding effect of inductance [13, 15]. For delay computation the I$^{st}$ order transfer function dominant pole is evaluated, because the dominant pole decides the delay of a distributed network. Thus the equivalent resistance is given as

$$r = 0.65R_s + 0.36Z_o + R_T \tag{1}$$

Where $R_s$ is the source resistance and $Z_o$ is the characteristic impedance ($Z_0 = \sqrt{L_T/C_T}$) and $R_T$, $C_T$ and $L_T$ represents total line resistance, capacitance and inductance discussed in table 1 for any length of line.

## III. FORMULATION OF BIT LINE DELAY FOR CURRENT MODE SIGNALING IN LONG INTERCONNECT

All the applications of electronics and electrical system those contains digital integrated IC's with memory unit. The speed enhancement of on-chip memories/systems is an important area of research, which is targeted in this work.

### A. Problem Definition

In this work a bit line delay is modeled when a read operation performed on CMOS SRAM.

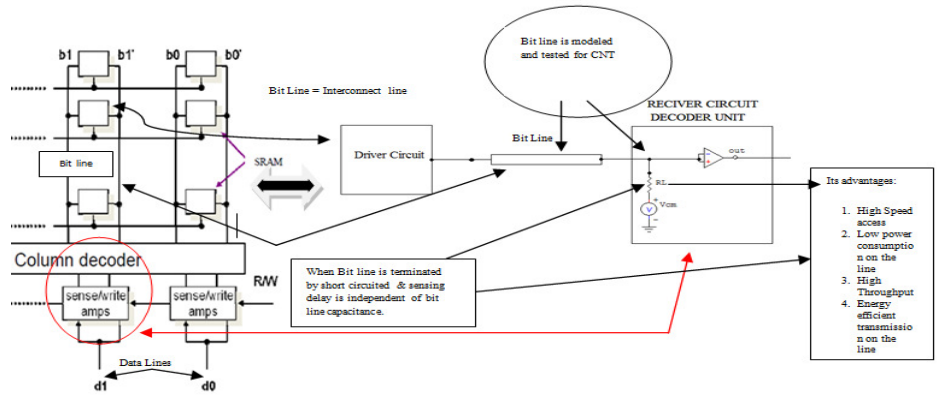

Fig.1 Equivalent circuit model for the proposed problem (when SRAM drives a bit line)

Current mode signaling technique is exploited for fast access/transfer of information to data line of any microprocessor/microcontroller. For current mode signaling a system consist of a driver circuitry, interconnect line and followed by receiver circuitry having a decoding unit. The problem targeted in this work detailed in figure 1.

### B.   Mathematical formulation

In this paper, a case of SRAM cell drives the long interconnect lines is approximated in problem definition section but it is approximated as inverter just for reducing the complexity level. as shown in figure 2.

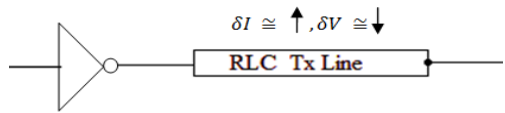

Fig.2 Long interconnect lines represented by distributed RLC transmission lines, and driven by an inverter

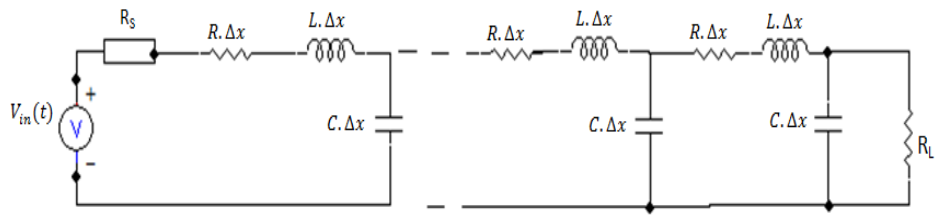

Fig. 3 Current mode interconnect model

And in the present work distributed RLC model for a current mode interconnects is shown in figure 3. Line parameters are designated   $R$, $L$, and $C$ as unit length resistance, inductance and capacitance respectively, $\Delta x$ is the length of each lumped section and $R_S$ is the source resistance. It is very much clear from literature that current mode signaling differs from voltage mode in that interconnect terminates at a finite resistance in addition to capacitive load. In this work delay model is proposed for resistive load. As shown in figure 3, the principle of current mode signaling is that by loading the line with finite impedance, the dominant pole of the system shifts, results in a smaller time constant and thus less delay.

Long transmission line is modeled as a linear time invariant distributed network. Furthermore, to represent a constant current and voltage on the line the differential equations representation is used, where voltage $V(x,t)$ and $V((x + \Delta x), t)$ and current $I(x,t)$ and $I((x + \Delta x), t)$, are represented at the source and load ends *at* $x = 0$ and $x = d$ (length of line) respectively. Figure 4 shows the equivalent distributed *rc* interconnect model. Here $r$ represent unit length equivalent resistance and $c$ represent unit length capacitance of the interconnect.

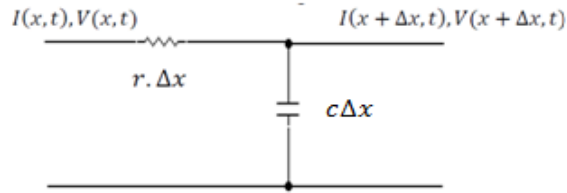

Fig.4 Distributed *rc* interconnect line model.

For Constant current

$$\frac{V(x,t) - V(x + \Delta x, t)}{r.\Delta x} = I(x,t)$$

For $\Delta x \to 0$

$$I(x,t) = -\frac{1}{r}\frac{\partial V(x,t)}{\partial x} \qquad (2)$$

For Constant Voltage

$$I(x,t) - I(x + \Delta x, t) = c\Delta x \frac{\partial V(x,t)}{\partial t}$$

For $\Delta x \to 0$

$$\frac{\partial I(x,t)}{\partial x} = -c\frac{\partial V(x,t)}{\partial t} \qquad (3)$$

Substituting (2) into (3), reduces to

$$\frac{\partial^2 V(x,t)}{\partial x^2} = rc \frac{\partial V(x,t)}{\partial t}$$

Thus $\quad \frac{\partial^2 V(x,t)}{\partial x^2} - rc \frac{\partial V(x,t)}{\partial t} = 0 \qquad (4)$

s- domain representation of (4) is

$$\frac{\partial^2 V(x,s)}{\partial x^2} - rcs\, V(x,s) = 0 \qquad (5)$$

Figure 5 gives the *rc* distributed model of an interconnect line. $V_{in}(t)$ is the time varying input signal, and $R_S$ is a source resistance with $R_L$ resistive load. $r.\Delta x$ and $c.\Delta x$ represent small increments in the value of unit length resistance and capacitance down the interconnect line.

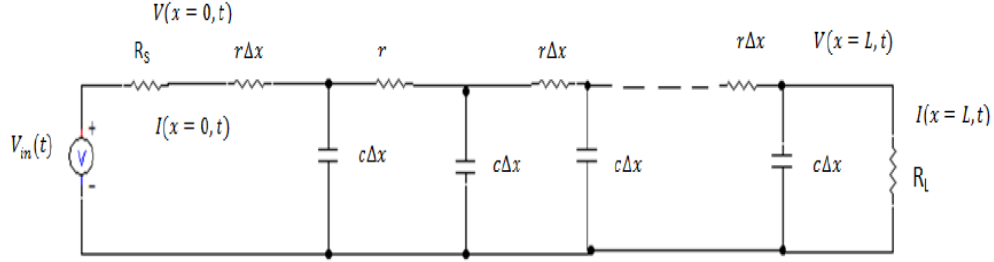

Fig. 5 Interconnect line modeled as a distributed line.

The solution of partial differential equation (5) in terms of voltage and current on the line is given by

$$V(x,s) = A_{11} \sinh(\sqrt{scr}\, x) + B_{11} \cosh(\sqrt{scr}\, x) \qquad (6)$$

$$I(x,s) = -\sqrt{\frac{sc}{r}}\, [A_{11} \cosh(\sqrt{scr}\, x) + B_{11} \sinh(\sqrt{scr}\, x)] \qquad (7)$$

Applying the boundary conditions on (6) and (7), $A_{11}$, $B_{11}$ with $R_L$ as resistive load termination are obtained. And the boundary conditions are:

$$V_{in}(s) = V(x = 0, s) + I(x = 0, s) R_S$$

$$V(x = d, s) = I(x = d, s) R_L$$

$$A_{11} = -\frac{V_{in}(s)[\cosh(\sqrt{scr}\, d) + \sqrt{\frac{sc}{r}}\, R_L\, \sinh(\sqrt{scr}\, d)]}{\left(\frac{sCR_L R_S}{r} + 1\right) \sinh(\sqrt{scr}\, d) + \sqrt{\frac{sc}{r}}\, (R_L + R_S) \cosh(\sqrt{scr}\, d)}$$

$$B_{11} = \frac{V_{in}(s)[\sinh(\sqrt{scr}\, d) + \sqrt{\frac{sc}{r}}\, R_L\, \cosh(\sqrt{scr}\, d)]}{\left(\frac{sCR_L R_S}{r} + 1\right) \sinh(\sqrt{scr}\, d) + \sqrt{\frac{sc}{r}}\, (R_L + R_S) \cosh(\sqrt{scr}\, d)}$$

Using the values of $A_{11}$ and $B_{11}$ at the load end (6) and (7) reduce to (8).

$$\frac{V(x = d, s)}{V_{in}(s)} = \frac{\sqrt{\frac{sc}{r}}\, R_L}{\left(\frac{sCR_L R_S}{r} + 1\right) \sinh(\sqrt{scr}\, d) + \sqrt{\frac{sc}{r}}\, (R_L + R_S) \cosh(\sqrt{scr}\, d)}$$

(8)

Let $\sqrt{scrd} = u$

This leads to

$$\frac{V(x=d,s)}{V_{in}(s)} = \frac{\frac{u}{dr} R_L}{\left(1 + \frac{u^2}{r^2 d^2} R_L R_S\right) \sinh(u) + \frac{u}{dr}(R_L + R_S)\cosh(u)}$$

$$\frac{V(x=d,s)}{V_{in}(s)} = \frac{1}{\left(\frac{rd}{uR_L} + \frac{u}{rd} R_S\right)\left(\frac{e^u - e^{-u}}{2}\right) + \left(1 + \frac{R_S}{R_L}\right)\left(\frac{e^u + e^{-u}}{2}\right)}$$

(9)

Rewriting (8) as:

$$f(u) = \left(\frac{a}{u} + bu\right)\left(\frac{e^u - e^{-u}}{2}\right) + c\left(\frac{e^u + e^{-u}}{2}\right)$$

(10)

where, $a = \frac{rd}{R_L} = \frac{R_1}{R_L}$, $b = \frac{R_S}{rd} = \frac{R_S}{R_1}$, $c = \left(1 + \frac{R_S}{R_L}\right)$

On simplifying equation (10), it gives:

$$f(u) = 1/2\left[e^u\left(\frac{a}{u} + bu + c\right) + \left(c - bu - \frac{a}{u}\right)e^{-u}\right]$$

(11)

By solving (11), $f(u)$ finally reduces to

$$f(u) = (a+c) + \left(b + \frac{c}{2!} + \frac{a}{3!}\right)u^2 + \left(\frac{b}{3!} + \frac{c}{4!} + \frac{a}{5!}\right)u^4 + \left(\frac{b}{5!} + \frac{c}{6!} + \frac{a}{7!}\right)u^6 + \ldots$$

(12)

Substituting the value of

$$u = \sqrt{scr}\, d$$

$cd = C_1$, Total capacitance of interconnect line of length 'd'.

$rd = R_1$, Total effective resistance of interconnect line of length 'd', and (12) becomes:

$$f(u) = (a+c) + \left(b + \frac{c}{2!} + \frac{a}{3!}\right)C_1 R_1 s + \left(\frac{b}{3!} + \frac{c}{4!} + \frac{a}{5!}\right)(C_1 R_1)^2 s^2 + \left(\frac{b}{5!} + \frac{c}{6!} + \frac{a}{7!}\right)(C_1 R_1)^3 s^3 + \ldots$$

(13)

Thereby the distributed network is further approximated to a first order transfer function as shown below where $a_1$ is the dominant pole that determines the delay of the line.

$$\frac{V(x=d,s)}{V_{in}(s)} = \frac{1}{(a+c) + \left(b + \frac{c}{2!} + \frac{a}{3!}\right) C_1 R_1 s}$$

(14)

First order transfer function is equivalent to

$$\frac{V(x=d,s)}{V_{in}(s)} = \frac{K_1}{s + a_1} = \frac{V_{dd}}{s}\left(\frac{K_1}{s + a_1}\right)$$

(15)

Finally, system response is converted into time domain and gives:

$$V(x = d, t) = \frac{V_{dd}}{a + c}[1 - e^{-a_1 t}]u(t) \tag{16}$$

Hence the delay time is computed as:

$$\tau_d = \frac{1}{a_1} = \frac{\left(b + \frac{c}{2!} + \frac{a}{3!}\right)}{a + c} C_1 R_1 \tag{17}$$

Substituting the values of value of $a$, $b$, and $c$, (17) reduces to:

$$\tau_d = \frac{[\frac{R_S}{R_1} + \frac{1}{2}\left(1 + \frac{R_S}{R_L}\right) + \frac{1}{6}\frac{R_1}{R_L}]C_1 R_1}{[\frac{R_1}{R_L} + \left(1 + \frac{R_S}{R_L}\right)]} \tag{18}$$

The delay of the proposed system will be equal to (18) in which $R_S$, $R_1$, $C_1$ and $R_L$ represents a source resistance, total line effective resistance, total line capacitance and Load resistance respectively. Further this model is validated analytically & by performing simulations for different length of interconnect with carbon nanotube (CNT) type of materials as interconnect and results are calculated for various interconnect hierarchy and comparison with existing model is also presented.

## C. Damping Factor

For current mode signalling, a lumped system model can be used for the approximate evaluation of the line inductance effect .This analysis of an *RLC* transmission line is compared to the analysis of a lumped *RLC* circuit [20].

The interconnect is modelled as a single section RLC circuit with $R_T$= R.d, $L_T$= L.d, $C_T$= C.d as shown in figure 6.

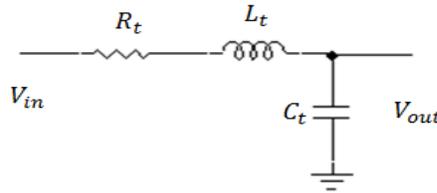

Fig. 6 Simple lumped *RLC* circuit model of an Interconnect line.

The poles of the circuit are

$$P_{1,2} = \omega_0[-\xi \pm \sqrt{(\xi^2 - 1)}] \tag{19}$$

and the damping factor $\xi$ is

$$\xi = \frac{R \cdot d}{2} \sqrt{\frac{C}{L}}$$

(20)

As (20) implies, if $\xi$ is greater than one, the poles are real and the effect of the inductance on the circuit is small. The greater the value of $\xi$, the more accurate the *rc* model become.

On the other hand, as $\xi$ become less than one, the poles become complex and oscillation occur. In that case, the inductance cannot be neglected. This relationship is physically intuitive, since $\xi$ represents the degree of attenuation the wave suffers as it propagates a distance equal to the length of the line. As this attenuation increase, the effect of the reflections decrease and the *rc* model becomes more accurate. Therefore $\xi$ is useful figure of merit that anticipates the importance of considering in a particular interconnect line.

## IV. RESULTS AND DISCUSSION

As the process technology downscales, smaller devices and wires come into picture. Also now the performance and reliability issues matter more. The downscaling of technology makes the devices faster but the wires get slow. The downscaling of wires increases their resistivity due to surface roughness, grain boundary scattering and further, due to higher current densities, electromigration problem becomes a headache. The wires' parameters have to be taken account of to analyze the performance of the chip. So the circuit parameters i.e. resistance, capacitance and inductance are necessarily to be analyzed before introducing any new material as interconnect in the VLSI chips. Carbon Nanotube Interconnect Analyzer (CNIA) [21, 22 & 23] has been used as the simulating tool for CNT bundle interconnect. CNIA simulator has four windows for varying inputs named as Geometry, Process, CNT and ambient. The interconnect dimensions that have been considered in proportion to that provided by the PTM are given in table 1 at 45nm.

Table 1 Simulation Parameters- Interconnect dimensions

| Parameters →Technology node | | Width | Thickness | Spacing | Height | Dielectric const |
|---|---|---|---|---|---|---|
| 45nm | Local | 68nm | 136nm | 68nm | 136nm | 2.1 |
| | Intermediate | 95nm | 240nm | 95nm | 136nm | 2.1 |
| | Global | 310nm | 820nm | 310nm | 136nm | 2.1 |

The various values of resistance, capacitance and inductance derived using the CNIA simulation tools have been used and the results derived are detailed in table 2. Their implications are also given. And table 2 presents the simulation and analytical results delay analysis of current mode interconnects. Firstly, measurements for the delay have been done at 45nm technology node. The gamma type network of transmission lines is used as an equivalent circuit to represent the interconnects in the T-Spice tool of Tanner EDA Inc [19]. With increase in the length of interconnect the total resistance of the line increases & magnitude of line delay increases. In table 2 the present work at CNT type of material provides a significant reduction in line delay, when compared to [15] for local, intermediate & global interconnect average reduction factor is 30%, 23% and 23% respectively, whereas when present work is tested at aluminium (Al) & Copper (Cu) material results are found to be close agreement to each other. These results are estimated when no condition of deep submicron regime is applied on the interconnects. It means that gate delay is greater than line delay and during calculation of (18) no assumption has been taken. Further it is also verified by simulation results shown in table 3. There is always probability of increase in error when data is transmitted over long length interconnects. Whereas same error becomes 0.68% for global length of interconnect. Consequently, it is justified that the proposed model is applicable for global length interconnects. The average error between the analytical and simulated results remains 2.2% for local, intermediate and global length of interconnect. So it is better at global length of interconnect. Figure.7 shows the variation of delay with interconnects length. It is seen that delay increases with length of interconnect. This is in accordance with the analytical model given by (18).

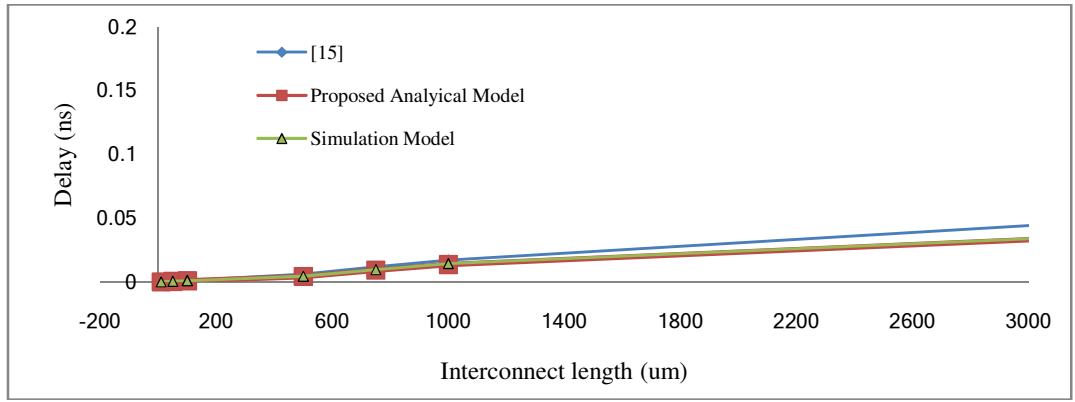

Fig. 7 Variation of delay with interconnects length for current mode

A very good agreement is seen between the analytical and SPICE simulation results. From figure 7 it is seen that delay obtained is lesser in case of proposed RLC interconnect when it is compared to [15]. This improvement in delay factor is because of moment approximation method is used in [15] while the proposed model overcomes the approximation.

Further the inductive effect is more prominent at lower technology node is presented. The delay contribution due to self- inductance could be significant in this case needs to be considered. A line 10mm long has an inductance of about 19.37nH, at 180nm node when combined with a line resistance of 220Ω, and a source resistance of 2.5kΩ, the short circuit time constant is about 7ps. Therefore the delay due to the line inductance will be negligible. This is finally limited by the wave propagation delay, which is in the order of 102ps for 10mm long line when calculated for 180nm node. The rise of delay due to inductance is because inductance does not allow sudden change of current on line. The delay introduced by inductance (table 4) when analyzed at 45nm technology node the results are completely different. The inductance effect in case of CNT materials cannot be neglected because inductance is dominating i.e 129.129nH for 10mm long line. So, the line propagation delay is equal to line inductance delay. Throughput comparison of proposed models is shown in figure. 8. It is decreasing with the length of interconnect. From this throughput can be predicted for different length of interconnects.

Electrical and Electronics Engineering: An International Journal (ELELIJ) Vol 3, No 4, November 2014

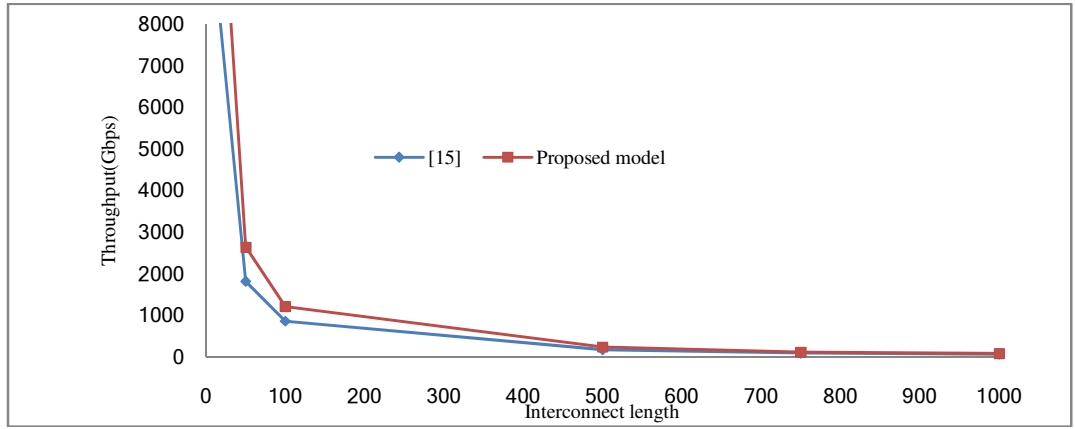

Fig. 8 Throughput variation with the length of interconnects for CNT material.

This all become possible due to change in signalling technique Throughput energy product can be taken as figure of merit for current mode technique shown in figure 8. And energy dissipated in transmission of a single bit across the interconnect is

$$E_{bit} = \frac{1}{2}(C_{int})V_{dd}^2 \qquad (21)$$

Where $E_{bit}$ represent energy per bit, $V_{dd}$ is the supply voltage, $C_{int}$ is the equivalent capacitance of the interconnect. The worst case power dissipated per line is the product of the throughput and bit energy for a periodic pulse train. From figure 8 it is concluded that throughput energy product of proposed is reduced to 66.71% when compared to voltage mode of interconnect & a comparison between voltage mode and current mode is presented in table 4 with the condition of DSM means line delay is greater than gate delay. After assumption the equation in (18) will reduce to $\tau_{VM} = \frac{R_1 C_1}{2}$ for voltage mode (VM) and for current mode (CM) it will reduce to $\tau_{CM} = \frac{R_1 C_1}{6}$. It means that CM dissipates 0.015pJ energy where as VM consume 0.045pJ for a single bit transmission across the interconnect over CNT material. So, the current mode interconnect system is an energy efficient data transmission system. First important aspect of this modelling is that when bit line is terminated by short circuit at load end. Then it is found that current mode signalling is superior to voltage mode signalling with the condition of DSM on interconnects. The superiority factor of current mode is 66.66% over voltage mode interconnects shown in table 4 and graphically presented in figure 9 at the end.

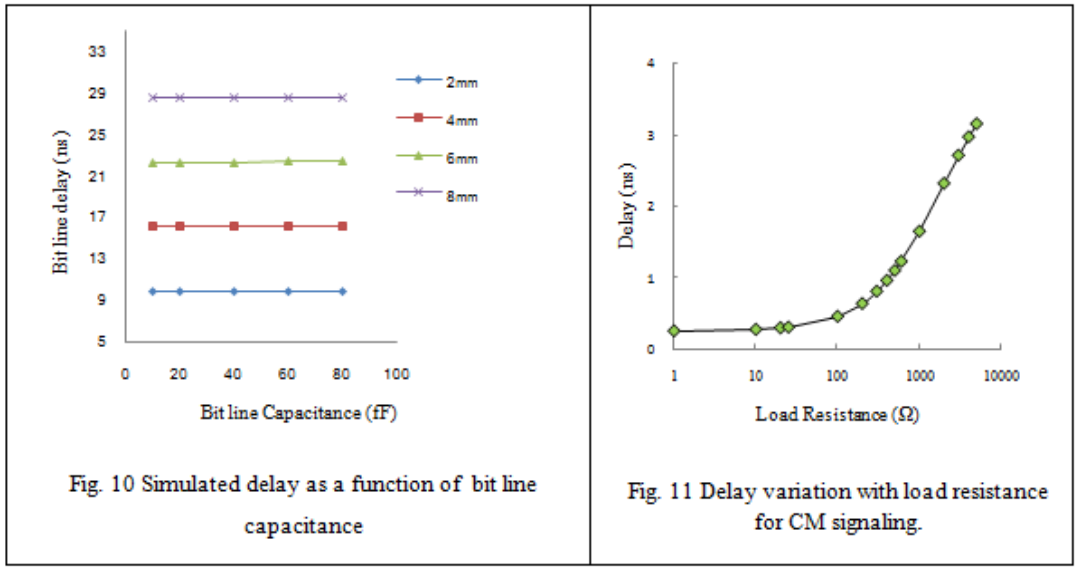

Fig. 10 Simulated delay as a function of bit line capacitance

Fig. 11 Delay variation with load resistance for CM signaling.

Second important aspect of short circuit is that it will reduce the dynamic power consumption on the line because of reduced voltage swing on interconnecting line. And reduced swing on the line can be estimated using the step response analysis [17]. Third important aspect which we concluded that due to short circuit the bit line delay is found to be independent of bit line capacitance by simulation it is investigated as shown in figure 10. For a metal line of 2mm, 4mm, 6mm, 8mm and 10mm long, the value of $R_T$, $C_T$, is calculated using PTM [18] and for different values bit line load capacitance i.e. 10fF, 20fF, 40fF, 60fF, 80fF the line delay is found to be independent of bit line load capacitance. The simulation results shown in figure 10 confirm that line delay is insensitive for extremely small value of load capacitance whereas it increases for larger value of bit line load capacitance. Keeping in view the advantages shown by current mode signaling once line is short circuited at termination end. And line delay variation with the load resistance is shown in figure 11. It increases with the value of load resistance.

So in order to implement the proposed current mode technique, current transporting circuits are needed with a low input resistance. Current gain is not needed at this stage, a unity gain transfer is sufficient. Further circuitry can take care of any required amplification or conversion to voltage mode. The circuit function can be identified as the current conveyor. This has been defined as device having virtual short circuit input port, and a unity gain transfer characteristic from input to output. Further possible current conveyor circuit is shown which can be used at termination end shown below.

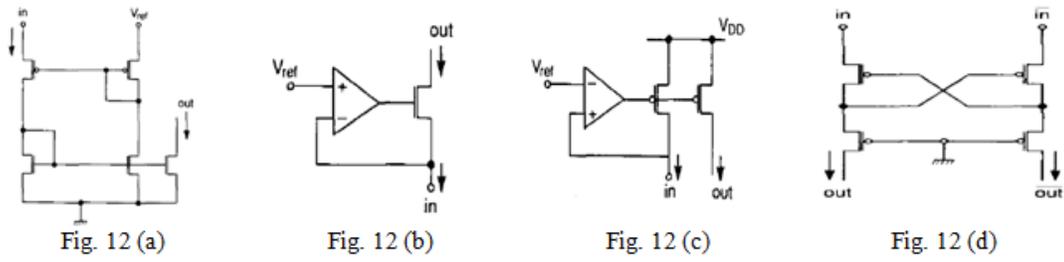

Fig. 12 (a)   Fig. 12 (b)   Fig. 12 (c)   Fig. 12 (d)

The circuit in figure 12 (a) [24] is based on unity current gain positive feedback. It keeps the input terminal at $V_{ref}$ by matching the currents in the p-channel transistors. The output current is equal to, or a multiple of, the input current, depending on the relative n-channel transistors sizes. The circuit in figure 12(b) [24] and (c) [25] are based on negative feedback. For the circuit in figure 12 (b) the current transfers is limited to unity, while for the circuit in figure 12 (c) current gain is possible. But these circuit options seem feasible for single interconnect lines. However, they have the drawback of complexity when applied in SRAM data path. Since two bit lines are involved, two current conveyers would be required, resulting in a relatively complex solution. A simpler approach, providing two virtually short-circuited current inputs, two current outputs, and requiring only four equal-sized transistors is indicated in figure 12(d). So this is a better solution for problem defined in figure 1 for SRAM cells. Thus it is analyzed that current mode signaling is superior to voltage mode signalling in various aspects.

## V. CONCLUSIONS

In the upcoming technology nodes, looking into the need for performance effective integrated circuits, CNT is a good option to replace Cu as an interconnect material.

In this paper a novel analytical delay model for current mode signaling is developed and presented. By using this proposed model dominant pole is computed from the first order system function. It is analyzed for different current mode circuit parameters to determine the nature of current mode circuits. It is also observed that the proposed model for CNT type of material, global and intermediate length of interconnects provides average reduction 23% whereas at Al and Cu results are in close agreement with existing model. Further with the length of interconnect the simulation results deviates the analytical model at global length of interconnect by 0.68%. So it is better to use CNT material at global length of interconnects. When the proposed model results evaluated at 45nm node, it is found that CNT provides 81.78% reduction in delay and w.r.t. to Al it is further reduced upto 86.80%. In DSM mode the superiority factor between current versus voltage remains 66.66%, once load is shorted at termination end. Throughput, Bit line delay and energy consumed during bit transmission is also discussed and presented. By simulative investigation, it is found that bit line delay is insensitive to extremely small value of load capacitance. Various topologies for sensing the signal at receiver termination is also discussed. And finally, it is concluded that the use of current mode techniques can lead to significant speed enhancement in long VLSI interconnects. This proposed current mode technique can significantly impact chip access times and architecture trade-offs for future fast CMOS SRAM design. Current mode signal receivers can be used to significantly reduce the line delays in CMOS VLSI chips. Secondly, figure of merit have been developed that determine the relative accuracy of a *rc* model on-chip interconnects. The derived expression along with accuracy analysis can serve as a convenient tool for delay estimation with minimal computation during design.

## Acknowledgement

The authors acknowledge with gratitude the technical and financial support from YMCA University of Science & technology, Faridabad, Haryana, India, for providing EDA tool facilities in Electronics Circuit Design and simulation Lab and National Council Of YMCAs of India, Govt of Haryana, India and the Central Agencies for Development Aid, Bonn, Germany for technical support.

Table 2 Comparison between present and existing work at 45nm. (without DSM condition)

| Interconnect Lengths (μm) | Interconnects Delay for 45nm technology Node ||||||  Reduction Improvement in Present Work (CNT) than Ref. [15] (CNT) % |
| | Present Work ||| Ref. [15] ||| |
| | Al Delay (ns) | Cu Delay (ns) | CNT Delay (ns) | Al Delay (ns) | Cu Delay (ns) | CNT Delay (ns) | |
| --- | --- | --- | --- | --- | --- | --- | --- |
| 10 | 0.000045 | 0.000037 | 0.00007 | 0.000056 | 0.000048 | 0.000104 | 32.69 |
| 50 | 0.000706 | 0.000516 | 0.00038 | 0.000745 | 0.000566 | 0.000553 | 31.28 |
| 100 | 0.002498 | 0.001791 | 0.000831 | 0.0025 | 0.001851 | 0.001173 | 29.16 |
| 500 | 0.026074 | 0.018665 | 0.004289 | 0.024829 | 0.018271 | 0.00616 | 30.37 |
| 750 | 0.054243 | 0.039149 | 0.009142 | 0.050383 | 0.037296 | 0.011771 | 22.33 |
| 1000 | 0.090636 | 0.065906 | 0.013714 | 0.082889 | 0.06166 | 0.017111 | 19.85 |
| 5000 | 0.286767 | 0.206207 | 0.052564 | 0.274987 | 0.203215 | 0.071167 | 26.14 |
| 7500 | 0.601816 | 0.432954 | 0.089358 | 0.56218 | 0.414952 | 0.116602 | 23.36 |
| 10000 | 1.008322 | 0.730307 | 0.133034 | 0.926656 | 0.687107 | 0.168439 | 21.02 |

Table 3 Material based comparison at 45nm.

| | Interconnect Lengths (μm) | Material based comparison of proposed model at 45nm ||||||  |
| | | Present Work ||| Simulation CNT (ns) | % Error Col.4 v/s Col. 5 | % Reduction using Proposed model CNT v/s Cu | % Reduction using Proposed model CNT v/s AL |
| | | Al Delay (ns) | Cu Delay (ns) | CNT Delay (ns) | | | | |
| --- | --- | --- | --- | --- | --- | --- | --- | --- |
| Local | 10 | 0.000045 | 0.000037 | 0.00007 | 0.00007 | 0 | 89.189 | 55.555 |
| | 50 | 0.000706 | 0.000516 | 0.00038 | 0.000392 | 3.1 | 26.356 | 46.175 |
| | 100 | 0.002498 | 0.001791 | 0.000831 | 0.000859 | 3.3 | 53.601 | 66.733 |
| Int. | 500 | 0.026074 | 0.018665 | 0.004289 | 0.00442 | 3.0 | 77.021 | 83.550 |
| | 750 | 0.054243 | 0.039149 | 0.009142 | 0.009515 | 4.0 | 76.648 | 83.146 |
| | 1000 | 0.090636 | 0.065906 | 0.013714 | 0.01438 | 4.8 | 79.191 | 84.869 |
| Global | 5000 | 0.286767 | 0.206207 | 0.052564 | 0.053204 | 1.2 | 74.509 | 81.670 |
| | 7500 | 0.601816 | 0.432954 | 0.089358 | 0.090336 | 1.0 | 79.360 | 85.151 |
| | 10000 | 1.008322 | 0.730307 | 0.133034 | 0.1327 | 0.2 | 81.783 | 86.806 |

Table 4 Comparison between voltage and current mode interconnect delay using CNT material at 45nm in DSM when $\tau_{int} > \tau_{gate}$ then $R_{int} > R_S$ or $R_1 > R_S$.

| Length Interconnect (μm) | | CM delay (ps) $R_L=0$ | VM delay (ps) $R_L=\infty$ | % of Reduction Col.2 v/s Col.3 |
|---|---|---|---|---|
| Local | 10 | 0.0238 | 0.071 | 66.48 |
| | 50 | 0.130 | 0.391 | 66.75 |
| | 100 | 0.286 | 0.858 | 66.67 |
| Intermediate | 500 | 1.472 | 4.418 | 66.68 |
| | 750 | 3.171 | 9.514 | 66.67 |
| | 1000 | 4.798 | 14.394 | 66.67 |
| Global | 5000 | 17.999 | 53.999 | 66.67 |
| | 7500 | 31.030 | 93.090 | 66.67 |
| | 10000 | 46.245 | 138.737 | 66.67 |

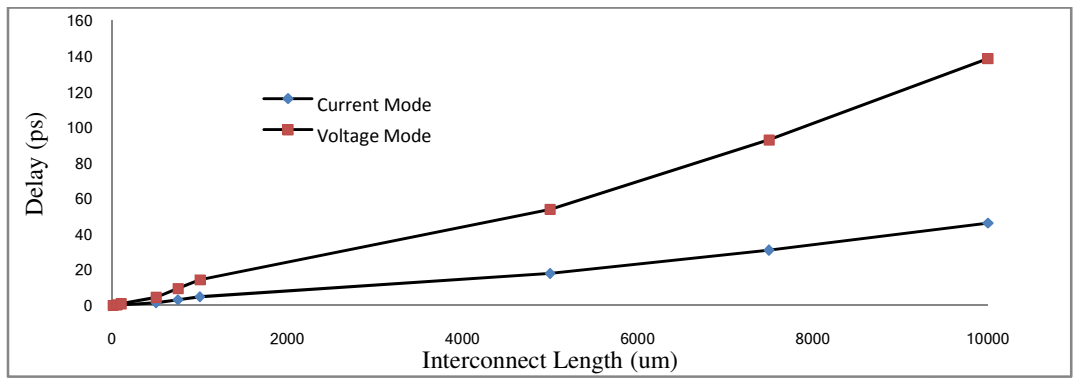

Fig. 9 Comparison between voltage and current mode interconnect using CNT material

**Authors**

**Sunil** Jadav received the B.Tech. degree in Electronics & Communication Engineering from the Guru Jambheshwar University of Science & Technology, Hissar Haryana, India, in 2007 and the M.Tech degree in VLSI Design & Automation from National Institute of Technology, Hamirpur, Himachal Pradesh, India in 2010. In 2011, he joined YMCA University of Science & Technology, Faridabad, Haryana, India (State University), as an Assistant Professor in Electronics Engineering Department. Currently he is also pursuing Ph.D. degree in Electronics Engineering Department of YMCA University of Science & Technology Faridabad. 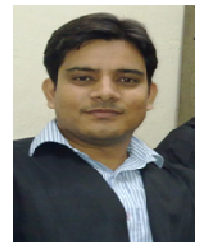
Before joining YMCAUST, he also worked as Assistant Professor in Electronics & Communication Engineering Department of National Institute of Technology, Hamirpur. During his work in N.I.T Hamirpur, he has effectively utilized and worked on various VLSI CAD Tools/Semiconductor Process and on field programmable gate array architecture development and low-power circuit design. He has published more than 20 papers in refereed journals and Conferences. His research interests include analog IC design/CAD with particular emphasis in low-power electronics for portable computing and wireless communications, and High Speed Low power VLSI Interconnect.

**Dr. Munish Vashishath** received his B.Tech in Electronics and Telecommunication Engineering from North Maharashtra University, Jalgaon in the year 1997, M.E in Electronics and Control Engineering with Hons. from Birla Institute of Technology and Science, Pilani in 2000 and Ph.D (Semiconductor Devices) from Thapar University, Patiala in the year 2010. From Dec 2007, he is serving as Associate Professor in Electronics Engg. at YMCA University of Science & Technology, Faridabad. His research interest includes Microelectronics, Semiconductor Devices Modeling & 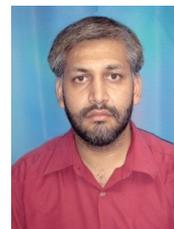 Simulation and VLSI Technology. Before 2007, he served many reputed institutes like Thapar University, NIT, Kurukshetra and NIT, Hamirpur. He has also published 40 papers in reputed Journals and Conferences.

**Dr.(Mrs.) Rajeevan Chandel has done** pre engineering (Gold Medalist) in 1986 from D.A.V College Kangra, H.P university and after that she has received her B.Tech in Electronics and Communication Engineering from Thapar Institute of Engineering & Technology, Patiala in the year 1990, M.Tech in Integrated Electronics Engineering from Indian Institute of Technology, Delhi, India in 1997 and Ph.D (VLSI Design, Microelectronics) from Indian Institute of Technology, Roorkee, Utrakhand, India in the year 2005. From Dec 1997, she is serving as 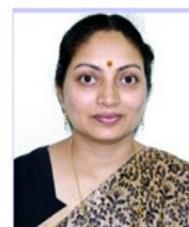 Professor & Head in Electronics & Communication Engg. at N.I.T Hamirpur Himachal Pradesh, India. Her research interest includes Microelectronics, Semiconductor Devices Modeling & Simulation and Low-Power VLSI Circuit and Interconnect Design. She has also published more than 140 papers in reputed Journals and Conferences.